\documentclass[aps,prb,twocolumn,superscriptaddress]{revtex4-2}

\usepackage{amsmath, amssymb,  graphicx}

\usepackage[colorlinks=true ,urlcolor=blue,urlbordercolor={0 1 1}]{hyperref}

\usepackage{color}
\usepackage{xcolor}
\usepackage{braket}

\usepackage[utf8]{inputenc}
\usepackage{bm}
\usepackage{verbatim}
\usepackage{graphicx}
\usepackage{amsmath}
\usepackage{color}
\usepackage{float}
\usepackage{bbm}
\usepackage{amssymb}
\usepackage{slashed}
\usepackage{wasysym,bm,bbm,dsfont}

\begin{document}

\title{Continuous transition from Fermi liquid to A fractional Chern insulator}

\author{Ya-Hui Zhang}
\affiliation{Department of Physics and Astronomy, Johns Hopkins University, Baltimore, Maryland 21218, USA}

\date{\today}

\begin{abstract}
    Recent experiments in moir\'e materials have observed fractional Chern insulators (FCI) at zero magnetic field, providing an opportunity to study the transition from FCI to the more conventional phases such as Fermi liquid (FL) and superconductor (SC) by tuning the interaction strength or bandwidth.   In this work, we formulate a critical theory for a continuous transition at the filling $\nu=\frac{2}{3}$ between the FL and a FCI* phase that hosts an additional neutral sector, but has the same transport signatures as the usual FCI.  In our framework, this corresponds to a transition from a composite Fermi liquid (CFL) to a superfluid* phase of the composite bosons. The Fermi liquid close to the transition has an additional factor $|z_i-z_j|^6$ in its wavefunction. Also, the transport behavior at high temperature on the FL side is actually like an `anyon gas' phase on top of the FCI,  with $\rho_{xy}$ close to $\frac{3}{2}\frac{h}{e^2}$. FL behavior with $\rho_{xy} \approx 0$ is recovered only at very low temperature. We also briefly discuss the possibility of a chiral superconductor as the descendant of this strongly correlated FL and the potential relevance to  the twisted MoTe$_2$ system.
\end{abstract}

\maketitle

\section{Introduction}

Fractional Chern insulators\cite{sun2011nearly,sheng2011fractional,neupert2011fractional,wang2011fractional,tang2011high,regnault2011fractional,bergholtz2013topological,parameswaran2013fractional,PhysRevB.84.165107}  at fractional filling of  the topological  bands of moir\'e systems\cite{zhang2019nearly,wu2019topological} have attracted lots of attentions after the experimental realizations  in twisted MoTe$_2$ homobilayer \cite{cai2023signatures,zeng2023integer,park2023observation,PhysRevX.13.031037} and in rhombohedrally stacked multilayer graphene proximated to an aligned hexagon boron nitride (hBN) \cite{lu2024fractional}. There are already many theoretical studies \cite{yu2020giant,devakul2021magic,li2021spontaneous,crepel2023fci,wang2023fractional,reddy2023fractional,2023arXiv230809697X,2023arXiv230914429Y,PhysRevLett.131.136501,PhysRevLett.131.136502,morales2023magic,song2024phase,dong2023theory,zhou2023fractional,dong2023anomalous,guo2023theory,kwan2023moir} trying to understand the ideal condition to optimize FCIs and their competing states. From a purely conceptual perspective, the currently realized FCI is likely in the same phase as the fractional quantum Hall (FQH) state at high magnetic field\cite{stormer1999fractional,stormer1999nobel}, which has been well understood. One natural next question is: can we realize completely new physics unprecedented in the Landau level systems at high magnetic field?  One unique opportunity of the moir\'e systems is that it is easy to tune the ratio between the bandwidth and the interaction. Therefore it is possible to study the evolution from the FCI to a conventional state such as a Fermi liquid or a superconductor.  Theories already exist for continuous transition between the bosonic Laughlin state and superfluid at filling $\nu=\frac{1}{2}$\cite{barkeshli2014continuous,song2023deconfined,lu2025continuous,wang2025emergent,lotrivc2025paired} and between a composite Fermi liquid (CFL) and Fermi liquid for electron at $\nu=\frac{1}{2}$\cite{barkeshli2012continuous,song2024phase}. In this work we turn our attention to electron at the filling $\nu=\frac{1}{3}$. A continuous transition between a FCI and a charge density wave (CDW) insulator has already been pointed out\cite{song2024phase}. However, the most robust state at larger bandwidth is the Fermi liquid (FL) and it is not clear whether the FL can directly transits into a FCI insulator.

In this work, we provide a theory of a possible continuous transition from the FL to an intermeidate FCI phase (dubbed as FCI*) at $\nu=\frac{2}{3}$ by decreasing the bandwidth. The FCI* phase has exactly the same resistivity tensor $\rho$ as the familiar FCI, but it hosts an additional neutral sector which only contributes to thermodynamic measurement (such as specific heat) and thermal transport. Unfortunately we are still not able to find a direct transition from FL to the conventional FCI without the additional neutral degrees of freedom. 

We use the composite boson theory, which can be constructed by the usual slave boson parton decomposition: $c(\mathbf x)=b(\mathbf x) f(\mathbf x)$. Then we put $f$ in the FCI phase. Then the composite boson $b$ feels an effective flux of $\frac{2\pi}{3}$ per unit cell. If $b$ is in a composite Fermi liquid (CFL) phase, electron is in a conventional FL phase, but with an additional factor $\prod_{i<j}|z_i-z_j|^6$ in its wavefunction. On the other hand, if $b$ forms a superfluid (SF), electron is in the usual FCI phase. In this language, the FL-FCI transition corresponds to the CFL-SF transition of the composite boson $b$. The later is still a quite challenging problem. Here we make a compromise and consider a simpler transition for the boson. Starting from the CFL of $b$, we simply condense the boson operator $b$ and reach a SF* phase with superfluid of the boson coexisting with a neutral sector. For the electron $c$, this is the FCI* phase. Therefore the critical theory of the FL-FCI* transition is from the condensation of a boson field. In the FCI* phase, the composite boson $b$ is in a mixture of condensation part and a normal part formed by a composite fermion $\psi$.   We expect the FCI* phase to occupy at most a very narrow region of the phase diagram and the density of $\psi$ may quickly vanish at a second critical point between the FCI* phase and the usual FCI phase by further decreasing the bandwidth.

The existence of an intermediate FCI* phase may depend on model details and the critical point may be replaced with another competing state such as a CDW insulator. On the other hand, some of our predictions may only rely on a correlated FL phase proximate to the FCI, as long as it hosts a small gap to the composte boson excitations. We find that the resistivity tensor has a strong deviation from typical FL phase when increasing the temperature above the gap of composite boson. At finite $T$, there is a finite density of thermally excited composite bosons, which depletes the density associated with the Fermi surface. The high temperature regime has a $\rho_{xy}$ close to $\frac{3}{2} \frac{h}{e^2}$ and a small $\rho_{xx}$, similar to an `anyon gas' phase on top of a FCI. At lower temperature the FL behavior is recovered with $\rho_{xy}\approx 0$. In principle, this unusual temperature dependence does not rely on a critical point, and may be associated with a strongly correlated Fermi liquid with cheap composite boson excitations. We propose to search for such an unusual finite-temperature transport and possible lower temperature superconductor in twisted MoTe$_2$ at and close to $\nu=\frac{2}{3}$.

\section{Framework: composite boson construction}

We consider the filling $\nu=\frac{1}{m}$ with $m$
 an odd integer. The most relevant case is the filling of $\nu=\frac{1}{3}$, but we will keep $m$ as a generic odd integer in the following.  Our theory may also apply to $\nu=1$ to describe a transition from FL to a quantum anomalous Hall(QAH)* insulator.

 We hope to work in the langauge of composite boson. Technically the flux attachment can be done by a parton construction of the electron operator $c$ as: $c(\mathbf r)=b(\mathbf r) f(\mathbf r)$, where $b$ is the composite boson and $f$ is a fermion. We will always put $f$ in a quantum Hall state at the filling $\nu=\frac{1}{m}$. This construction introduces an internal $U(1)$ gauge field $a_\mu$ shared by $b$ and $f$. In terms of the wavefunction, the electron wavefunction is in the form

 \begin{equation}
     \psi(z_1,...,z_N)=\psi_b(z_1,...,z_N)\prod_{i<j}(z_i-z_j)^m
 \end{equation}
 where the second factor is the wavefunction of $f$ in the Laughlin state at the filling $\frac{1}{m}$.   Now one can view $b$ as a composite boson from attaching $m$ fluxes to the electron.  In the conventional Landau level, the attached flux cancels the external flux and the composite boson $b$ simply condenses, leading to a fractional quantum Hall state. Now we consider a problem with a lattice at zero magnetic field. In this case, the external flux can be viewed as $0$ or $2\pi$ per unit cell, which are actually indistinguishable.  If there is an external $2\pi$ flux per unit cell, then again we can simply let the boson $b$ condense and obtain a FCI state. On the other hand, we may view the external flux to be $0$ per unit cell. In this case the composite boson $b$ actually feels an effective flux of $-2\pi$ per unit cell.  If the dispersion of $b$ is flat, then it may be itself in a quantum Hall state instead of a superfluid phase. Effectively, the boson $b$ is at filling $\nu_{m}=-\frac{1}{m}$ per flux and can form a composite Fermi liquid (CFL) phase. As we will see later, this leads to a conventional Fermi liquid (FL) of the original electron. Therefore the FL to FCI transition corresponds to the CFL to superfluid (SF) transition of the composite boson $b$.

A low energy effective theory of the electron is described by the following Lagrangian:

\begin{equation}
        \mathcal L_c=\mathcal L_b[b,a]+\mathcal L_f[f,A-a]
    \label{eq:total_theory}
\end{equation}
where $a_\mu$ is the internal U(1) gauge field. $A_\mu$ is the external probing field. We have assigned the physical charge to $f$. In this work, $f$ is always kept in the corresponding Laughlin state, described by the effective theory

 \begin{equation}
     \mathcal L_{f}[A-a]=-\frac{m}{4\pi}\beta d \beta+\frac{1}{2\pi}(A-a) d \beta
     \label{eq:theory_f_m}
 \end{equation}
 where $\beta_\mu$ is another internal U(1) gauge field introduced to capture the FQH state of $f$. $\alpha d \beta$ is the abbreviation of the Chern-Simons term $\epsilon_{\mu \nu \sigma}\alpha_\mu \partial_\nu \beta_\sigma$ with $\epsilon$ the anti-symmetric tensor with convention $\epsilon_{012}=1$.

\subsection{FL}

The composite boson is at effective filling $\nu=-\frac{1}{m}$ and can form a CFL phase. Then its wavefunction is in the form

\begin{equation}
    \psi_b(z_1,...,z_N)=\psi_{\mathrm{FS}}(z_1,...,z_N)\prod_{i<j}(z_i^*-z_j^*)^m
\end{equation}
where $\psi_{\mathrm{FS}}$ is the Slater determinant for a Fermi surface state.

In the end, the total wavefunction for electron is:

\begin{equation}
     \psi_{\mathrm{FL}}(z_1,...,z_N)=\psi_{\mathrm{FS}}(z_1,...,z_N)\prod_{i<j}|z_i-z_j|^{2m}
\end{equation}

This  corresponds to a correlated FL phase. The extra factor $\prod_{i<j}|z_i-z_j|^{2m}$ can reduce inter-particle repulsion.

In field theory, the CFL phase of the composite boson is described by the standard Halperin-Lee-Read (HLR) theory\cite{halperin1993theory}:

\begin{equation}
    \mathcal L_b[a]=\mathcal L_{\mathrm{FS}}[\psi,\alpha]+\frac{m}{4\pi}\alpha_2 d \alpha_2+\frac{1}{2\pi}(a-\alpha)d\alpha_2
    \label{eq:cfl_boson_m}
\end{equation}
where $\mathcal L_{\mathrm{FS}}[\psi,\alpha]$ describes a Fermi surface formed by the fermion $\psi$ coupling to the gauge field $\alpha$:

\begin{align}
    \mathcal L_{\mathrm{FS}}[\psi,\alpha_1]&=\psi^\dagger(t,x)(i\partial_t+\alpha_{1;0}+\mu)\psi(t,x)\notag \\ 
    &~~~-\frac{\hbar^2}{2m_\psi} \psi^\dagger(t,x)(-i\partial_\mu+a_{1;\mu})^2 \psi(t,x)
\end{align}

 The above theory can be  derived by a secondary parton construction $b=\psi \psi_2$ and then we put $\psi_2$ in the $\nu=-\frac{1}{m}$ Laughlin state.  The internal U(1) gauge field $\alpha_2$ is introduced to encode the Laughlin state of $\psi_2$. $\alpha$ is the internal U(1) gauge field shared by $\psi$ and $\psi_2$.

Combining Eq.~\ref{eq:total_theory}, Eq.~\ref{eq:theory_f_m} and Eq.~\ref{eq:cfl_boson_m}, we can integrate $a$ to lock $\alpha_2=\beta$, then the Lagrangian for the electron is:

\begin{equation}
    \mathcal L_{c}=\mathcal L_{FS}[\psi,\alpha]+\frac{1}{2\pi}(A-\alpha)d \beta
\end{equation}
where the self Chern-Simons (CS) terms for $\beta$ are canceled.  Now we can integrate $\beta$, which locks $\alpha_\mu=A_\mu$. In the end the Lagrangian is:

\begin{equation}
    \mathcal L_{\mathrm{FL}}=\mathcal L_{\mathrm{FS}}[\psi,A]
\end{equation}
which describes a FL phase with $\psi \sim c$.

 \subsection{FCI}

The boson $b$ feels an effective magnetic flux of $-2\pi$ per unit cell. However, with a lattice, the composite boson $b$ can acquire a dispersion and form a superfluid phase instead of a CFL phase. In this case, we get a FCI phase for electron. In Eq.~\ref{eq:total_theory}, the condensation of $b$ simply higgses $a_\mu$.  After that, the electron is described by the following theory:

\begin{equation}
    \mathcal L_c=-\frac{m}{4\pi} \beta d \beta+\frac{1}{2\pi} a d \beta
\end{equation}
which is the standard effective theory for the FCI phase at the filling $\nu=\frac{1}{m}$.

\section{FL to FCI* transition}

In our framework, the FL to FCI transition can be formulated as the CFL to superfluid (SF) transition of the boson at the filling $\nu=\frac{1}{m}$. 
A theory of the CFL-SF transition is still quite challenging because it is very difficult to capture a continuous destruction of a Fermi surface. The same problem exists in the Fermi liquid to Mott insulator transition\cite{senthil2008theory}. The best we can do is to obtain a continuous transition between the CFL phase and an SF phase coexisting with a neutral sector (described by a doped CFL). This can be done by simply condensing the boson $b$ starting from the CFL phase.  In the original electron language, the other side is a FCI insulator, but with an additional neutral sector which only manifests in thermal measurement.  We dub this more exotic insulator as FCI*. In the following, we provide a critical theory for a potential continuous  FL-FCI* transition, tuned by bandwidth or interaction at the fixed density.

In the language of the composite boson $b$, we start from the CFL phase in Eq.~\ref{eq:cfl_boson_m} and then simply condense the boson $b$ to reach a superfluid phase with $\langle b \rangle \neq 0$.  In this intermediate phase, there are both quantum Hall component and superfluid component. Only at a second transition does the quantum Hall component completely disappear.   The transition is described by the following theory:

\begin{align}
    \mathcal L_b[a]&=\mathcal L_{\mathrm{FS}}[\psi,\alpha]+\frac{m}{4\pi}\alpha_2 d \alpha_2+\frac{1}{2\pi}(a-\alpha)d\alpha_2 \notag \\ 
    &+\varphi^*(i\partial_t+a_0)\varphi-\frac{1}{2m_\varphi}\varphi^*(\vec \nabla-i \vec a)\varphi-g|\varphi|^4-s|\varphi|^2
    \label{eq:qcp_boson}
\end{align}
where $\varphi$ is a complex field to capture the condensation transition of $b$, so it couples to the same gauge field $a_\mu$. Note that there is generically a linear derivative of the time for $\varphi$. The transition is tuned by $s$. When $s>0$, we have $\varphi$ gapped and we recover the theory of the CFL for the composite boson $b$.  Correspondingly, the electron is in the FL phase. When $s<0$, we have $\varphi \neq 0$ which higgses $a$.  This turns out to describe the FCI* phase in electron language.

\subsection{FCI* phase}

On the $s<0$ side, we condense $\varphi$ and $a_\mu$ is higgsed. For the composite boson with density $n_b=\frac{1}{m}$, now we have a two-fluid picture: a superfluid part with density $n_\varphi=|\langle \varphi \rangle|^2$ and a neutral part with density $n_\psi=n_b-n_\varphi$. The composite boson is now in a superfluid phase coexisting with a dark sector which hosts a quantum-Hall like phase at effective filling $n_{\psi}=\frac{1}{m}-|\langle \varphi \rangle|^2$.  When increasing $|s|$, $n_\psi$ gradually decreases to zero, after which we have a pure superfluid phase. In the intermediate regime, the neutral sector of the composite boson is described by the Lagrangian:

\begin{equation}
    \mathcal L_b=\mathcal L_{\mathrm{FS}}[\psi,\alpha]-\frac{1}{2\pi} \alpha d \alpha_2+\frac{m}{4\pi } \alpha_2 d \alpha_2
\end{equation}
with the density $n_\psi=\frac{1}{m}-n_\varphi$ with $n_\varphi=|\langle \varphi \rangle|^2$. 

We can also integrate $\alpha_2$ and obtain

\begin{equation}
    \mathcal L_b=\mathcal L_{\mathrm{FS}}[\psi,\alpha]-\frac{1}{4\pi m} \alpha d \alpha
\end{equation}

Due to the CS term, there is an average flux $\frac{1}{2\pi} \langle \vec \nabla \times \alpha \rangle= -m n_\varphi $. Thus $\psi$ feels an effective flux with filling $\nu=-\frac{\frac{1}{m}-n_\varphi}{m n_\varphi}$. As a result the Fermi surface is replaced with Jain sequences, as in the usual bosonic quantum Hall system.  In our case this sector does not couple to the physical electro-magnetic probes and can be detected only by thermodynamic measurement and thermal transport measurement.

Now we return to the electron picture.  We  combine Eq.~\ref{eq:theory_f_m} to reach the full theory of the electron:

\begin{align}
    \mathcal L_{\mathrm{FQAH*}}=\mathcal L_{\mathrm{FS}}[\psi,\alpha]-\frac{1}{4\pi m}\alpha  d \alpha 
    -\frac{m}{4\pi} \beta d \beta+\frac{1}{2\pi} A d \beta
\end{align}
Note that $a_\mu$ is removed.  We now have two decoupled sectors. The first part is a `dark' sector with quantum-Hall like phase depending on the averange flux $\frac{1}{2\pi} \langle \vec \nabla \times \alpha \rangle= -m n_\varphi $ and the density $n_\psi=\frac{1}{m}-n_\varphi$. $n_\varphi$ is continuously varied inside the FCI* phase until $n_\psi$ vanishes and the `dark' sector disappears.   The second part is the standard theory of the FCI phase with $\sigma_{xy}=\frac{1}{m} \frac{e^2}{h}$.  If one only cares about charge transport measurements, this FCI* phase is exactly the same as the usual FCI insulator.

\subsection{Critical theory}

The critical theory of the FL-FCI* transition is a combination of Eq.~\ref{eq:qcp_boson} and Eq.~\ref{eq:theory_f_m}.  For simplicity, we will integrate $\alpha_2$. In the following, we are only interested in the low-energy property of the QCP, so we do not need to care too much about the quantization of the gauge field. Thus, we will also integrate $\beta$.  The final theory is in the form

\begin{align}
    \mathcal L_{\mathrm{FL-FCI^*}}&=\mathcal L_{\mathrm{FS}}[\psi,\alpha]-\frac{1}{4\pi m} \alpha d \alpha -\frac{1}{2\pi m}(A-\alpha) d a\notag \\
    &+\varphi^*(i\partial_t+a_0)\varphi-\frac{1}{2m_\varphi}\varphi^*(\vec \nabla-i \vec a)^2\varphi-s|\varphi|^2 \notag \\ 
    &-g|\varphi|^4 -\lambda(\varphi^\dagger (\psi \mathcal M_{\alpha}^m)+h.c.)+\frac{1}{4\pi m} A d A
    \label{eq:critical_FL_FCI*}
\end{align}

When $s>0$, $\varphi$ is gapped and can be ignored. Then integration of $a$ locks $A=\alpha$ and one recovers the FL phase with the Lagrangian to be simply $\mathcal L_{\mathrm{FS}}[\psi,A]$.  When $s<0$, $\varphi$ condenses and $a$ is higgsed.  Then we have a `dark' sector of $\psi$ decoupled from the FCI part of $\frac{1}{4\pi m} A d A$. Note that we have lost track of the gapped anyons  as a result of the integration of $\beta$. 

In the FCI* phase, we have the constraint that $n_\psi+n_{\varphi}=n=\frac{1}{m}$. Therefore the density of the neutral fermion $\psi$ gradually shrinks and eventually disappears.  We also have $\frac{\langle d \alpha\rangle}{2\pi}=-m n_{\varphi}$, so $\psi$ is at effective magnetic filling $\nu_\psi=-\frac{\frac{1}{m}-n_{\varphi}}{m n_{\varphi}}$.   Basically one can convert $\psi$ to the boson $\varphi$ by creating $-m$ fluxes of $\alpha$ at the same time. Once $\varphi$ is condensed, the density of $\psi$ is not conserved anymore, which is captured by the $\lambda$ term. There is likely a very rapid increase of $\varphi$ so the FCI* transits to the pure FCI phase by completely depleting the $\psi$ sector.

At QCP, we can  also identify the microscopic electron operator as $c\sim 
 \psi \mathcal M_a^{\dagger m}$, where $\mathcal M_a^\dagger$ creates a monopole of $a$.   From this equation, we can see that the electron $c$ and the critical boson $\varphi$ see each other as a vortex with charge $m$. In this sense, the condensation of $\varphi$ may be viewed as the proliferation of  the vortex of the electron, which destroys the coherence of the FL phase and leads to the FCI* phase. As we will see later, even within the FL phase, the vortex $\varphi$ may still be thermally excited.

\section{Transport property in the FL phase and the critical regime}

We are mainly interested in the transport property across the transition.  At zero temperature, the transport behavior is the same as a FCI phase for $s<0$ and the same as a FL phase for $s\geq 0$.  Even exactly at $s=0$,  the density of $\varphi$ is zero at $T=0$. At least in the  leading order, integration of $\varphi$ does not lead to any additional self energy for the gauge field $a_\mu$ for the $z=2$ theory. Therefore the systems behaves as if $\varphi$ is gapped and indistinguishable from a FL phase.

On the other hand, the resistivity tensor $\rho$ can strongly deviate  from FL at finite temperature even for $s>0$. At finite $T$, there is a  finite density of thermally excited $\varphi$ bosons.  But we need to fix $n_\psi+n_\varphi=\frac{1}{m}$ through a chemical potential term:

\begin{equation}
    \delta \mathcal L=+\delta \mu (\frac{1}{2\pi m} d\alpha+\varphi^\dagger \varphi)
\end{equation}

Note that the variation of $\alpha_0$ leads to $\frac{1}{2\pi m} d\alpha=n_\psi$.  Now the dispersion of $\varphi$ becomes $\epsilon(\mathbf k)=\frac{\hbar^2 k^2}{2m_\varphi}+\Delta$ with $\Delta=s-\delta \mu$.  Following the discussion in Sec.~\ref{append:phi_density_critical}, we expect a shift of the chemical potential $-\delta \mu= \frac{n_\varphi}{\kappa_{\mathcal M_\alpha}}$, where $\kappa_{\mathcal M_\alpha}$ is the susceptibility to generate a flux of $\frac{1}{2\pi m} d\alpha$, which is also proportional to the compressibility of the CFL phase of $b$.

In Sec.~\ref{append:phi_density_critical}, the expression of this compressibility is derived to be

\begin{equation}
    \kappa_{\mathcal M_{\alpha}}=\frac{6\pi}{m^2 W_\psi+3 W_\varphi(e^{\frac{\Delta}{k_B T}}-1)}
\end{equation}
where $W_{\varphi}=\frac{4\pi^2}{2m_\varphi a^2}$ and $W_{\psi}=\frac{4\pi^2}{2m_\psi a^2}$ have units of energy and may be viewed as the band width of $\varphi$ and $\psi$ respectively. $a=1$ is the lattice constant. One can see that $\kappa_{\mathcal M_\alpha}=0$ when $T=0$.

Even with a finite gap $\Delta=s-\delta \mu>0$, there can still be thermally excited bosons with density

\begin{equation}
    n_\varphi=\int \frac{d^2k}{(2\pi)^2} \frac{1}{e^{\frac{\frac{k^2}{2m_\varphi}+\Delta}{k_B T}}-1}=- \frac{2\pi T}{W_\varphi} \log(1-e^{-\frac{\Delta}{T}})
\end{equation}

\begin{figure}[ht]
    \centering
    \includegraphics[width=1\linewidth]{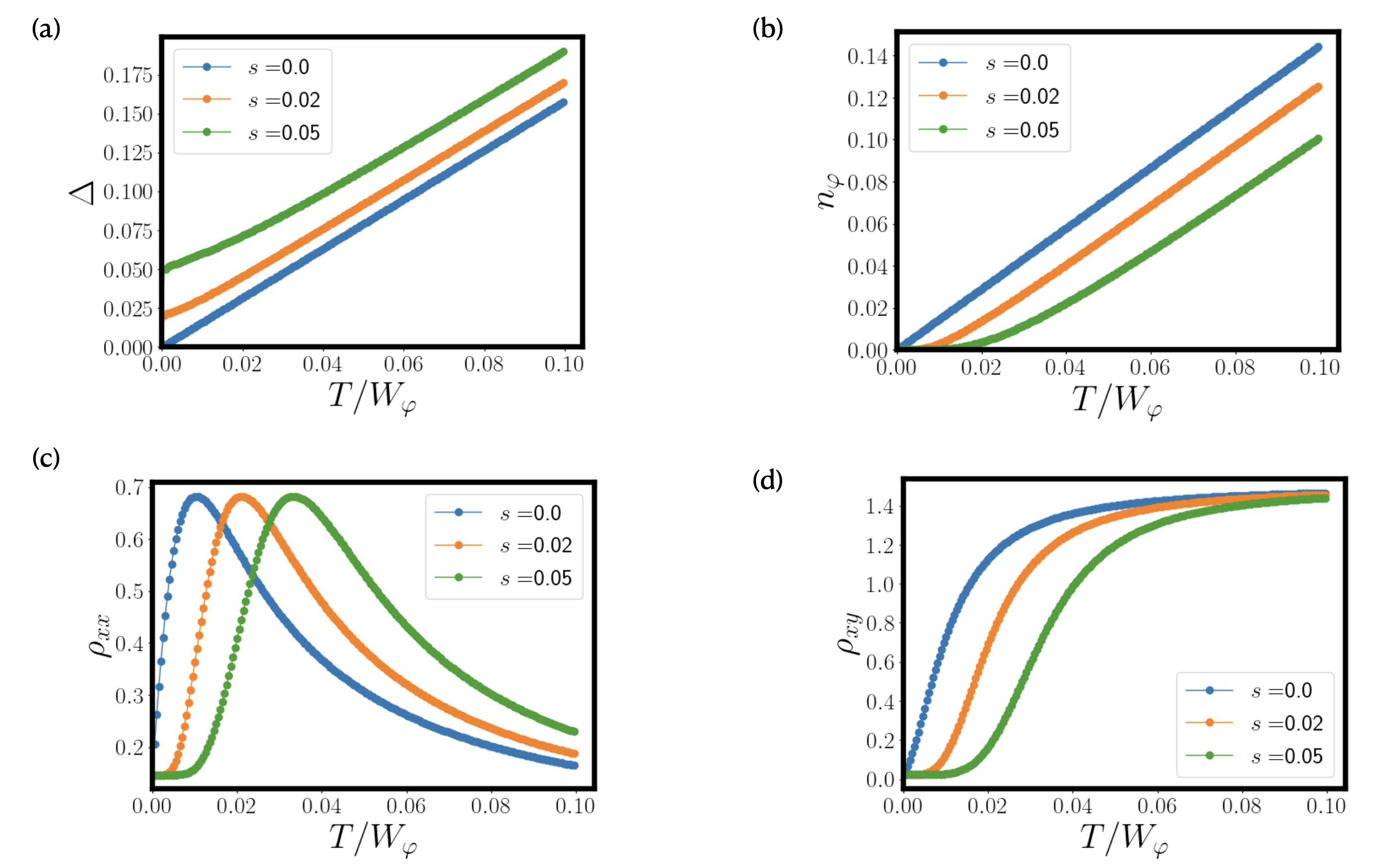}
    \caption{Resistivity $\rho_{xx}$ and $\rho_{xy}$ (in units of $\frac{h}{e^2}$) at the filling $\nu=\frac{2}{3}$, obtained from Eq.~\ref{eq:rho_c} and Eq.~\ref{eq:transport_calculation} using $W_\varphi=W_\psi=1$ and $\frac{h}{\tau_\varphi}=\frac{h}{\tau_\psi}=0.1 W_\varphi$. $s$ is also in units of $W_\varphi$. }
    \label{fig:transport_filling_2_3}
\end{figure}

Note here we ignore the interaction term $g$ for simplicity. For each $s\geq 0$, $\delta \mu$ and $n_\varphi$ can be self-consistently decided from the above two equations together with $\kappa_{\mathcal M_\alpha}(\Delta -s)=n_\varphi$. We plot the result in Fig.~\ref{fig:transport_filling_2_3} at filling $\nu=1-\frac{1}{3}$. At $T=0$, $\Delta=s$ as expected. But when $T>s$, we can see that $\Delta$ increases linearly with the temperature. Interestingly, $n_\varphi$ also increases linearly with $T$ when $T>s$ and can quickly reach $10\%$ of the total hole density which is $\frac{1}{3}$. This indicates that the FL behavior is only expected at very low temperature of $T<<s$. At higher temperature, a significant portion of the electrons are depleted from the Fermi surface and become the composite boson $\varphi$.

Let us calculate the conductivity tensor $\sigma$ for the filling $\nu=1-\frac{1}{3}$, assuming $\nu=1$ is a $C=-1$ Chern insulator. The conductivity tensor should be $\sigma=\sigma_h+\begin{pmatrix} 0 & -1 \\ 1 & 0 \end{pmatrix}$ in units of $\frac{e^2}{h}$, where the second part is from the  Chern insulator  at $\nu=1$. $\sigma_h$ is the conductivity tensor from the hole at the filling $\nu=\frac{1}{3}$.   The resistivity tensor $\rho_h=\sigma_h^{-1}$ can be obtained from the Ioffe-Larkin rule due to the parton construction $c=b f$: $\rho_h=\rho_b+\begin{pmatrix}0 & -3 \\ 3 &0 \end{pmatrix}$ in units of $\frac{h}{e^2}$, where the second part is $\rho_f$ from the Laughlin state of $f$.  $\rho_b=\sigma_b^{-1}$ and the conductivity $\sigma_b$ has two separate contributions from $\varphi$ and the doped CFL part. As derived in Appendix.~\ref{append:transport_critical}, we have:

\begin{equation}    \rho_h=\left[\sigma_{\varphi}I+\begin{pmatrix} \rho_{\psi}& \rho_{\psi;xy}+m \\ -(\rho_{\psi;xy}+m) & \rho_{\psi}\end{pmatrix}^{-1}\right]^{-1}+\begin{pmatrix}
        0 & -3 \\ +3 &0
    \end{pmatrix}
    \label{eq:rho_c}
\end{equation}
where $\sigma_\varphi$ is the conductivity from $\varphi$. $\rho_{\psi}$ is the longitudinal resistivity of $\psi$ and $\rho_{\psi;xy}$ is the Hall resistivity of $\psi$.

We can estimate the conductivity from $\varphi$ and $\psi$ from a simple relaxation-time approximation. As shown in Appendix.~\ref{append:transport_critical}, we have the expressions:

\begin{align}
    \sigma_{\varphi}&=\frac{2\pi \tau_\varphi}{h} k_B T \big(\frac{\Delta}{k_B T}-\log (e^{\frac{\Delta}{k_B T}}-1)\big) \notag \\ 
    \rho_{\psi;xx}&=  \frac{h}{\tau_\psi} \frac{1}{2 W_\psi n_\psi} \notag \\
    \rho_{\psi;xy}&=\frac{1}{n_\psi}-m
    \label{eq:transport_calculation_main}
\end{align}
where  $\sigma$ is in units of $\frac{e^2}{h}$ and $\rho$ is in units of $\frac{h}{e^2}$.  $n_\psi=\frac{1}{m}-n_\varphi$ and $n_\varphi$ can be obtained from the self consistent calculation described before. $\tau_\varphi$ and $\tau_\psi$ are relaxation time of $\varphi$ and $\psi$ respectively. Here we assume a temperature independent relaxation time, presumably from disorder.  We ignore the additional temperature dependence coming from interaction because the major temperature dependence come from $n_\varphi(T)$.

From Eq.~\ref{eq:transport_calculation_main}, we can see that $\sigma_\varphi=2\pi \frac{\tau_\varphi}{h} \Delta e^{-\frac{\Delta}{k_B T}}$ and is exponentially small close to $T=0$. When $T>>\Delta$, we find $\sigma_\varphi\sim \frac{2\pi \tau_\varphi}{h} k_B T \log \frac{k_B T}{\Delta}$ and increases linearly with the temperature.  At higher temperature, $\sigma_\varphi$ dominates and we find $\rho_h\sim \sigma_\varphi^{-1} I + \begin{pmatrix} 0 & -3 \\ 3 & 0 \end{pmatrix}$. Thus $\rho_{xx} \sim \frac{h}{2\pi \tau_\varphi} \frac{1}{k_B T}$ and $\rho_{xy} \approx \frac{3}{2}$ at high temperature, as shown in Fig.~\ref{fig:transport_filling_2_3}(c)(d).  The resistivity tensor is indeed similar to an `anyon gas' phase on top of a FCI phase with $\rho_{xy} \approx \frac{3}{2}$ when $T$ is larger than a few percent of $W_\varphi$, despite that we are on the FL side. The FL behavior with $\rho_{xy} \approx 0$ is recovered only at very low temperature $T<<s$. Note in reality there should be a $T^2$ dependence on $\rho_{xx}$ once the umklappa scattering is included.  At the QCP with $s=0$, we note that $\rho_{xx}$ first increases linearly with the temperature: $\rho_{xx} \sim \frac{A k_B T}{(A k_B T)^2+\frac{1}{9}}$ with $A \sim \frac{2\pi \tau_\varphi}{h}$.

\section{Discussion}

We have shown that the transport behaves strongly different from that of an FL phase at higher temperatures on the FL side of the transition. Although the transition itself depends on the existence of an intermediate FCI* phase, the unusual transport signature may exist on the FL side even if the FCI* and the transition are absent.  Within the FL phase, we may view $\varphi$ as a gapped excitation in addition to the usual excitations of the conventional FL.  The property of this strongly correlated FL is still described by Eq.~\ref{eq:critical_FL_FCI*}, but with a parameter $s>0$ characterizing the energy scale of the gapped excitation $\varphi$.

Because we have $c \sim \psi \mathcal M_a^{\dagger 3}$ (at filling $\nu=1-\frac{1}{3}$) and $\varphi$ has charge $1$ under $a_\mu$, $\varphi$ may be viewed as a $6\pi$ vortex of electron. Actually $\varphi$ is just the composite boson with an electron bound to $3$ fluxes. According to Eq.~\ref{eq:critical_FL_FCI*}, such a vortex excitation $\varphi$ has an energy gap $\Delta=s$ at $T=0$. When $T<<\Delta$, we have the usual FL behavior. But when $T>\Delta$, thermally excited composite bosons with a finite density $n_\varphi$ can dramatically alter the transport property. We conjecture that such a finite temperature deviation from FL transport behavior may exist in correlated metal phase close to the FCI phase, regardless of the nature of the transition.

The Fermi liquid phase has an additional factor $\prod_{i<j}|z_i-z_j|^6$ factor in its wavefunction, indicating strong correlation. It is known that the FL phase has a superconducting instability from the Kohn-Luttinger mechanism. Recent numerical simulation\cite{wang2025chiral} shows that the pairing instability is enhanced  close to the transition to the FCI phase.  It is interesting to study whether this additional factor in the wavefunction stimulates pairing. Even if a superconductor indeed emerges from the correlated FL phase, the transport measurements behave as a conventional metallic phase only in an intermediate temperature regime and is closer to an `anyon gas' phase at higher temperature.

\section{conclusion}

In summary, we develop a critical theory for a continuous quantum phase transition from a Fermi liquid (FL) to a fractional Chern insulator (FCI*) hosting an additional neutral sector, realized at fixed fillings $\nu=1/3$ and $2/3$. Within the FL phase, we identify a gapped composite boson excitation that profoundly impacts the finite-temperature resistivity tensor. In the high-temperature regime, thermal excitation of these bosons results in an ``anyon gas" phase, with the Hall resistivity approaching $\rho_{xy}\approx \frac{3}{2} \frac{h}{e^2}$ at $\nu=2/3$. Recent experiment observed a superconductor in twisted MoTe$_2$\cite{xu2025signatures} whose high temperature normal state has a large $\rho_{xy}$.  From our analysis, this unusual result may originate from  a  correlated FL with thermally excited composite boson excitations. Consequently, this provides a conventional alternative to the more exotic scenarios based on anyon superconductivity \cite{Laughlin1988PRL,Laughlin1988Science,HalperinAnyonSC1990,FisherLeeAnyonSC1991,tang2013superconductivity,kim2025topological,shi2024doping,darius2025doping}.

\section{Acknowledgement}

The work is supported by the National Science Foundation under Grant No. DMR-2237031.

\bibliography{refs}

\appendix 

\onecolumngrid

\section{Superfluid transition from a bosonic composite Fermi liquid\label{appendix:SF-CFL_boson}}

We consider a model with physical boson $b$ at density $\nu=\frac{1}{m}$ with odd $m$. We will discuss a CFL to SF* transition, where the SF phase coexists with a neutral sector. The transition is simply a condensation of $b$. SF* will transit to the conventional SF through a secondary confinement transition which we will not discuss here.

The critical theory is:

\begin{align}
    \mathcal L_{\mathrm{CFL-SF^*}}&=\mathcal L_{\mathrm{FS}}[\psi,\alpha]-\frac{1}{4\pi m} \alpha d \alpha +\frac{1}{2\pi m}A_b d \alpha-\frac{1}{4\pi m} A_b d A_b\notag \\
    &+\varphi^*(i\partial_t+A_{b;0})\varphi+\frac{1}{2m_\varphi}\varphi^*(\vec \nabla-i \vec A_b)^2\varphi-s|\varphi|^2 -g|\varphi|^4+\delta \mu(\frac{1}{2\pi m} d\alpha+|\varphi|^2) 
    \label{eq:SF*-CFL}
\end{align}
where $\vec A_b$ is the probing field for the physical boson.  $\alpha_\mu$ is an internal gauge field.  $\psi$ is the composite fermion from flux attachment of $m$ fluxes to the physical boson.  The chemical potential $\delta \mu$ is introduced to fix the total density $n_\psi+n_\varphi=\frac{1}{m}$. Note there is a relationship $\frac{d\alpha}{2\pi}=m \delta n_{\psi}=-m n_\varphi$ from the variation of $\alpha_0$. Here $d\alpha$ is an abbreviation of the magnetic flux $\nabla\times \vec \alpha$. The first line describes the CFL phase of the boson at filling $\nu=\frac{1}{m}$. The second line describes the condensation transition of the physical boson $\varphi \sim b$. Across the transition, the density $\varphi^2$ changes. Hence the transition is like a chemical potential tuned superfluid transition. Inside the SF* phase, we have a two fluid picture: condensation of the boson $b$ coexist with a doped CFL. In the SF* phase the neutral fermion $\psi$ also feels an effective flux $\frac{d\alpha}{2\pi}=-m \delta n_\psi$ locked to the change in density.

\subsection{Dilute boson gas at finite temperature in the CFL side}

Let us focus on the CFL phase in the side of $s>0$, where $n_{\varphi}=0, \frac{\langle d\alpha \rangle}{2\pi}=0$ at $T=0$.  However, at finite $T$, there are thermally excited bosons and the system is again described by a two fluid picture such that the CFL part is depleted by a finite $n_{\varphi}$.  We should view $s$ as the relative energy difference between the band bottom of  $\varphi$ and the Fermi energy of $\psi$. At $T=0$, we have $n_\varphi=0$ and thus $\delta \mu=0$. At finite $T$, we may expect a finite $-\delta \mu>0$. Let us define $\Delta=s-\delta \mu$, so the dispersion of $\varphi$ is $\epsilon_\varphi(\mathbf k)=\frac{\hbar^2 k^2}{2m_\varphi}+\Delta$.

The boson density can be estimated by a simple free-boson model (assuming $g=0$ for simplicity):

\begin{align}
    n_\varphi=\int \frac{d^2k}{(2\pi)^2} \frac{1}{e^{\frac{\frac{k^2}{2m_\varphi}+\Delta}{k_B T}}-1}=- \frac{2\pi T}{W_\varphi} \log(1-e^{-\frac{\Delta}{T}})
    \label{eq:density_phi}
\end{align}
 where we define $W_{\varphi}=\frac{\hbar^2}{2m_{\varphi}}(\frac{2\pi}{a})^2$, which can be viewed as the bandwidth of $\varphi$ in the scale of Brillouin zone (BZ). Note we used the unit $a=1$.
   We are interested in the regime of $\Delta,T<<W_\varphi$.  If $T<<\Delta$, $n_{\varphi}\approx \frac{2\pi T}{W_\varphi} e^{-\frac{\Delta}{T}}$ and is exponentially small. In this case we just  have the CFL phase and we expect $\delta \mu \approx 0$ and $\Delta \approx s$. On the other hand, in the regime of $T>>\Delta$, we have $n_{\varphi}\approx \frac{2\pi T}{W_{\varphi}}\log \frac{T}{\Delta}$.

   In the above we used $\Delta=s-\delta \mu$ instead of $s$. To compensate $n_\varphi$, we need to tune $\delta \mu$ to reduce $n_\psi$ by $\delta n_\psi=-n_\varphi$. It is easy to estimate $\delta \mu$ to be $\delta \mu=-\frac{n_\varphi}{\kappa_{\mathrm{CFL}}}$, where $\kappa_{\mathrm{CFL}}$ is the compressibility of the CFL. Note that we always have $\frac{d\alpha}{2\pi}=-m \delta n_\psi$, so the compresibility is associated with the diamagnetism $\chi_\psi=\frac{1}{12\pi m_\psi}$ of the $\psi$ fermion as: $\kappa_{\mathrm{CFL}}=(\frac{1}{2\pi m})^2\frac{1}{\chi_\psi}=\frac{3}{\pi m^2} m_\psi$.  Then $\delta \mu =- \frac{\pi m^2 
 n_{\varphi}}{3 m_\psi a^2}$, where $a=1$ is the lattice constant. Let us define $\frac{4\pi^2}{2 m_\psi a^2}=W_\psi$ as the bandwidth of $\psi$.  $\delta \mu$ can be obtained from the equation:

\begin{equation}
     \delta \mu=\frac{m^2}{3}\frac{ W_\psi}{ W_{\varphi}} T \log(1-e^{-(s+\delta \mu)/T})
\end{equation}

\begin{figure}
    \centering
    \includegraphics[width=0.95\linewidth]{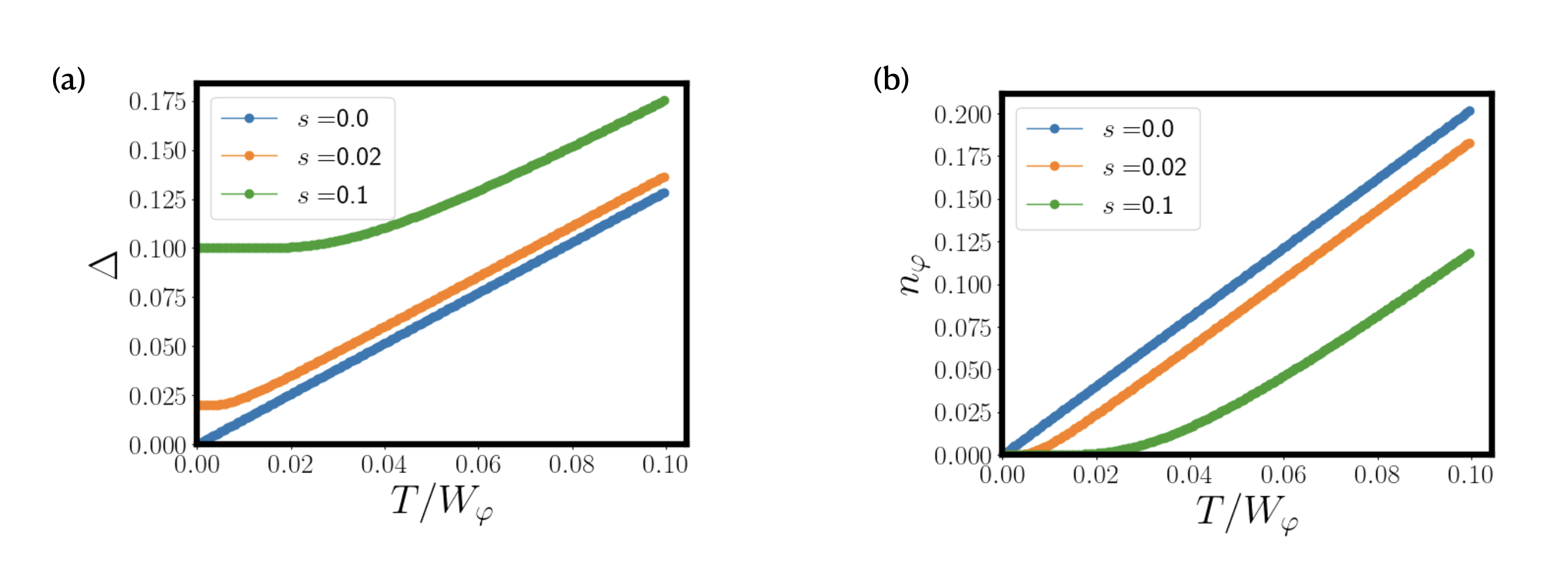}
    \caption{The dependence of $\Delta=s-\delta \mu$ and $n_{\varphi}$ on the temperature $T$ for different values of $s$ in units of $W_\varphi$.  We consider the case of $m=3$, so the transition is between the CFL and SF* phase at $\nu=\frac{1}{3}$.  We used $\frac{W_\psi}{W_\varphi}=\frac{4}{3}$ and $W_{\varphi}=1$.  (a) At $T=0$, $\Delta=s$. But at larger $T$, $\Delta$ is linear to the temperature due to the self adjusted $\delta \mu$. (b) For finite $s$, we have $n_\varphi \sim T e^{-\frac{s}{T}}$ at very low temperature. At slightly higher temperature, $n_\varphi$ is linear to $T$. For moederate $s$, a small temperature $T/W_{\varphi}=0.04$ can generate boson density at order of $0.1$. For $\nu=\frac{1}{3}$ filling, the density of $\psi$ is depleted by almost $30\%$. Therefore, at finite temperature, the property of the phase is very different from a CFL even at finite $s>0$.}
    \label{fig:density_T}
\end{figure}

We solved $-\delta_\mu$ and $n_\varphi$ self-consistently. We plot the gap $\Delta$ and $n_\varphi$ with temperature in Fig.~\ref{fig:density_T}. One can see that at finite $T$, $n_\varphi$ grows linearly with the temperature when $T>s$. For small $s>0$, $n_\varphi$ can reach $0.1$ even for a small temperature $T=0.04 W_\varphi$, suggesting that the behavior of the system deviates from the CFL phase at a moderate temperature, above which we must include a finite density of dilute bosons.

\subsection{Transport property}

In the following we discuss the transport property of this bosonic phase in the $s>0$ side.  At $T=0$, it is in the familiar CFL phase.   At finite $T$, we have contributions from both the doped CFL part and $\varphi$.  The conductivity of the physical boson is the sub of these two contributions:

\begin{equation}
    \sigma_b=\sigma_{\varphi}I+\begin{pmatrix} \rho_{\psi}& \rho_{\psi;xy}+m \\ -(\rho_{\psi;xy}+ m) & \rho_{\psi}\end{pmatrix}^{-1}
\end{equation}
where the second part is the conductivity from the CFL.

AT $T=0$, we expect $\sigma_{\varphi}=0$ simply because $n_{\varphi}=0$. Thus the transport at temperature regime $T<<\Delta$ just follows the familiar CFL phase. However, as shown in Fig.~\ref{fig:density_T}, $n_\varphi$ increases linearly with the temperature $T$ when $T>\Delta$. Hence a finite conductivity from the dilute boson $\varphi$ is expected in the `critical regime'.   It is known that the CFL part contributes a quite small $\sigma_{xx}$. So $\sigma_{xx}$ is dominated by $\sigma_\varphi$ at finite temperature.  In the following we perform a simple estimation of $n_{\varphi}$ and its temperature dependence. 

We use Botzmann equation with the relaxation time approximation:

\begin{equation}
    \partial_t f(\mathbf k, t)+ \frac{1}{\hbar}e\vec E \cdot \partial_{\mathbf k}  f(\mathbf k,t)=-\frac{1}{\tau} \left( f(\mathbf k, t)-f_0(\mathbf k,t)\right)
\end{equation}
where $f_0(\mathbf k,t)=\frac{1}{e^{\frac{\epsilon_k}{k_B T}}-1}$ is the Bose-Einstein distribution. $\epsilon_k=\frac{\hbar^2 k^2}{2m_\varphi}+\Delta$. $\tau$ is the relaxation time. Here we assume that the major source of the scattering is the disorder, so $\tau$ is momentum independent.    At linear order of $|\vec E|$, we can use $\partial_{\mathbf k} f(\mathbf k,t)=\partial_{\mathbf k} f_0(\mathbf k,t)=-\frac{1}{k_B T} \frac{\hbar^2 \mathbf k}{m_\varphi} \frac{e^{\frac{\epsilon_k}{k_B T}}}{\big(e^{\frac{\epsilon_k}{k_B T}}-1\big)^2}$. 

Define $\delta f(\mathbf k,t)=f(\mathbf k,t)-f_0(\mathbf k,t)$. In the DC limit, we ignore the $\partial_t$ term and obtain $\delta f(\mathbf k)= \frac{e\tau}{k_B T} \frac{\hbar \mathbf k \cdot \mathbf E }{m_\varphi} \frac{e^{\frac{\epsilon_k}{k_B T}}}{\big(e^{\frac{\epsilon_k}{k_B T}}-1\big)^2}$. Then the current is $\vec J=\int \frac{d^2 k}{(2\pi)^2} e \frac{\hbar \mathbf k}{m_\varphi} \delta f(\mathbf k)$. Then we find the conductivity:

\begin{align}
    \sigma^\varphi_{xx}(T)&=\frac{e^2\tau}{k_B T} \frac{\hbar^2}{m_{\varphi}^2} \int\frac{d^2k}{(2\pi)^2} k_x^2 \frac{e^{\frac{\epsilon_k}{k_B T}}}{\big(e^{\frac{\epsilon_k}{k_B T}}-1\big)^2}=\frac{e^2\tau}{k_B T} \frac{\hbar^2}{4\pi m_{\varphi}^2} \int_0^{+\infty} dk k^3 \frac{e^{\frac{\epsilon_k}{k_B T}}}{\big(e^{\frac{\epsilon_k}{k_B T}}-1\big)^2}=\frac{e^2\tau}{2\pi \hbar^2 k_B T} \int_0^{+\infty}d\epsilon_0 \epsilon_0 \frac{e^{\frac{\epsilon_0+\Delta}{k_B T}}}{(e^{\frac{\epsilon_0+\Delta}{k_B T}}-1)^2} \notag \\
    &=\frac{e^2}{h}\frac{\tau}{\hbar k_B T} \int_\Delta^{+\infty}d\epsilon (\epsilon-\Delta)\frac{e^{\frac{\epsilon}{k_B T}}}{(e^{\frac{\epsilon}{k_B T}}-1)^2}=\frac{e^2}{h} \frac{\tau}{\hbar} k_B T (\frac{\Delta}{k_B T}-\log(e^{\frac{\Delta}{k_B T}}-1))
\end{align}

When $k_B T<<\Delta$, we find that $\sigma^{\varphi}_{xx}(T)\approx \frac{\Delta \tau}{\hbar} e^{-\frac{\Delta}{k_B T}}$ in units of $\frac{e^2}{h}$ and is thus exponentially small at the $T=0$ limit.  On the other hand, in the critical regime with $k_B T>>\Delta$, $\sigma^\varphi_{xx}(T)\approx \frac{\tau}{\hbar} k_B T \log \frac{k_B T}{\Delta}$ in units of $\frac{e^2}{h}$. One can see that now we get a finite conductivity. With an AC field at frequency $\omega$, it is simply $\Delta^\varphi_{xx}\sim \frac{\tau}{1-i\omega \tau} \frac{k_B T}{\hbar} \log \frac{k_B T}{\Delta}$. Therefore there is a finite Drude weight which scales with the temperature as $D\sim \frac{k_B T}{\hbar} \log \frac{k_B T}{\Delta}$. Given that $\sigma_{xx}$ from the CFL part is vanishing, the conductivity of the physical boson is dominated by $\sigma_{xx}^{\varphi}$ and thus also has the Drude weight which scales as $T \log T$.  We conclude that there can be a finite Drude weight at finite temperature in the $s>0$ side even if the Drude weight vanishes at $T=0$.

\section{Physical property near the FL-FCI* transition}

Here we study the property in the critical regime of the FL-FCI* transition using the Eq.~\ref{eq:critical_FL_FCI*}.  But there is also a simpler picture. We can use the construction $c=b f$ and the put $f$ in the FCI state with $\sigma_{xy}=\frac{1}{m}$.  Then we consider a superfluid* to CFL transition of the boson $b$ discussed in Sec.~\ref{appendix:SF-CFL_boson}.  The resulting phase for the physical electron is exactly FCI* to FL transition.  Hence here we will quote some result in Sec.~\ref{appendix:SF-CFL_boson}.

In Eq.~\ref{eq:critical_FL_FCI*}, at the QCP with $s=0$, we can identify the microscopic electron operator as $c\sim 
 \psi \mathcal M_a^{\dagger m}$, where $\mathcal M_a^\dagger$ creates a monopole of $a$. When $s<0$, $a_\mu$ is Higgsed and $\psi$ should be viewed as a neutral fermion. In terms of physical electro-magnetic response, the system behaves as a FCI. When $s>0$, $\varphi$ is gapped at $T=0$. Then $\alpha$ is locked to $A$ and we get a Fermi liquid.  The focus below is to discuss the property at finite temperature, especially in the critical regime with $T>s$.

 For  the regime of $s>0$, the sector of $\varphi$ is similar to Eq.~\ref{eq:SF*-CFL} for the SF* to CFL transition for a boson at filling $\nu=\frac{1}{m}$. But there is one crucial difference, the probing field $A_b$ in Eq.~\ref{eq:SF*-CFL} now becomes a dynamical gauge field $a_\mu$. At $T=0$, because $\varphi$ is gapped, we can integrate $a_\mu$, which locks $\alpha_\mu=A_\mu$. Therefore both $a$ and $\alpha$ disappear together in the low energy and we get only a FL phase. At $T>0$, there may be a finite density of $\varphi$ and we need to discuss the effect of $a_\mu$.

\subsection{Density of $\varphi$ at finite temperature\label{append:phi_density_critical}}

At finite $T$, there may be  finite density of thermally excited $\varphi$ bosons.  But we need to fix $n_\psi+n_\varphi=\frac{1}{m}$ through a chemical potential term:

\begin{equation}
    \delta \mathcal L=+\delta \mu (\frac{1}{2\pi m} d\alpha+\varphi^\dagger \varphi)
\end{equation}

Note that the variation of $\alpha_0$ leads to $\frac{1}{2\pi m} d\alpha=n_\psi$.  Now the dispersion of $\varphi$ becomes $\epsilon(\mathbf k)=\frac{\hbar^2 k^2}{2m_\varphi}+\Delta$ with $\Delta=s-\delta \mu$.  Following the discussion in Sec.~\ref{appendix:SF-CFL_boson}, we expect a shift of the chemical potential $-\delta \mu= \frac{n_\varphi}{\kappa_{\mathcal M_\alpha}}$, where $\kappa_{\mathcal M_\alpha}$ is the susceptibility to generate a flux of $\frac{1}{2\pi m} d\alpha$ and is decided by the effective action of $\vec \alpha_\perp$ at finite momentum $\mathbf q$. It has two contributions: the diamagnetism of the fermion $\psi$ and an additional term from the integration of $a_0$.

Let us focus on the $\omega=0$ limit and try to get the finite $\mathbf q$ term for the self energy of the gauge field. At finite $T$, by integrating $\varphi$ and $\psi$, we get an effective action for $a_\mu$ and $\alpha_\mu$.

\begin{align}
    \mathcal L_{\mathrm{eff}}&= \frac{1}{2}\sum_{\omega,q} \chi_\psi q_x^2 |\mathbf \alpha_{y}(0,q_x)|^2-\frac{1}{2} \sum_{\omega,\mathbf q}\big(\frac{i q_x}{2\pi m} \alpha_y(0,\mathbf q) a_0(0,-\mathbf q) +\frac{-i q_x}{2\pi m} a_0(0,\mathbf q) \alpha_y(0,-\mathbf q)\big)+\frac{1}{2}\kappa_{\varphi}  | a_0|^2
\end{align}
where $\chi_\psi=\frac{1}{12\pi m_\psi}=\frac{W_\psi}{24\pi^3}$ (we set the lattice constant $a=1$), where $W_\psi=\frac{4\pi^2}{2m_\psi a^2}$ is the bandwidth of $\psi$. $\alpha_\perp(0,\mathbf q)$ is the transverse component of the gauge field perpendicular to the direction of $\mathbf q$. $\kappa_\varphi$ is the compressibility from $\varphi$ and can be evaluated from $-\frac{\partial n_\varphi}{\partial \Delta}$ using Eq.~\ref{eq:density_phi} to be: $\kappa_\varphi=\frac{2\pi}{W_\varphi} \frac{1}{e^{\frac{\Delta}{k_BT}}-1}$. In the above Lagrangian we ignore other terms because in the end we only care about the term for $|\alpha_y|^2$. Now we integrate $a_0$ and get the effective action for $\alpha_y$ to be:

\begin{equation}
    \mathcal L_{\mathrm{eff}}=\frac{1}{2}\sum_{\omega, \mathbf q} (\chi_\psi +\frac{1}{4\pi^2 m^2 \kappa_\varphi}) q_x^2 |\alpha_y(0,\mathbf q)|^2
\end{equation}

Then we get the compressibility to add a flux $d\alpha$ to be:

\begin{equation}
    \kappa_{\mathcal M_{\alpha}}=(\frac{1}{2\pi m})^2  \frac{1}{\chi_\psi +\frac{1}{4\pi^2 m^2 \kappa_\varphi}}=\frac{6\pi}{m^2 W_\psi+3 W_\varphi(e^{\frac{\Delta}{k_B T}}-1)}
\end{equation}

Clearly, at small $T$, $\kappa_{\mathcal M_{\alpha}}$ is exponentially small, consistent with the picture that $\alpha_\mu$ is higgsed.   $\Delta$ and $n_\varphi$ can be decided self-consistently by equation $\kappa_{\mathcal M_\alpha} (-\delta \mu)=n_\varphi$, or equivalently

\begin{equation}
    \frac{6\pi}{m^2 W_\psi+3 W_\varphi(e^{\frac{\Delta}{k_B T}}-1)}(\Delta -s)=- \frac{2\pi T}{W_\varphi} \log(1-e^{-\frac{\Delta}{T}})=n_\varphi 
    \label{eq:self_consistent_n_phi}
\end{equation}

\subsection{Transport in the critical regime\label{append:transport_critical}}

We now move to calculate the resistivity tensor, especially in the regime $T>s$ on the side of $s>0$. We will see that the resistivity behaves like an FL only when $T<s$ and becomes more like an anyon gas at a higher temperature.

We still use the slave boson construction $c=b f$. Then we put $f$ in a FCI state with $\sigma_{xy}=\frac{1}{m}$ in units of $\frac{e^2}{h}$. The FCI*-FL transition corresponds to the SF*-CFL transition of the boson $b$, which is captured by Eq.~\ref{eq:SF*-CFL}.  From the Ioffe-Larkin rule\cite{ioffe1989gapless}, we get the resistivity tensor of the physical electron:

\begin{equation}
    \rho_c=\left[\sigma_{\varphi}I+\begin{pmatrix} \rho_{\psi}& \rho_{\psi;xy}+m \\ -(\rho_{\psi;xy}+m) & \rho_{\psi}\end{pmatrix}^{-1}\right]^{-1}+\begin{pmatrix}
        0 & -m \\ +m &0
    \end{pmatrix}
    \label{eq:rho_c}
\end{equation}
where $\sigma_{\varphi}$ is the conductivity of the critical boson $\varphi$. $\rho_{\psi}=\rho_{\psi;xx}$ and $\rho_{\psi;xy}$ are the longitudinal and Hall resistivity of the fermion $\psi$.  When $T=0$, $\sigma_\varphi=0$ and one can easily find $\rho_c=\rho_\psi$, consistent with a Fermi liquid. On the other hand, when $T>>s$, $\sigma_\varphi$ is large and we expect $\rho_c=\sigma^{-1}_\varphi I_{2\times 2}+\begin{pmatrix}
        0 & -m \\ +m &0
    \end{pmatrix}$. Therefore the longitudinal resistivity is dominated by the critical boson, while $\rho_{xy}\approx -m \frac{e^2}{h}$, like an anyon gas on top of the FCI state.

We are interested in the evolution from the FL regime to the anyon-gas-like regime when increasing $T$.  For the purpose of illustration, we use the following simple model for the critical boson and the fermion $\psi$:

\begin{align}
    \sigma_{\varphi}&=\frac{e^2}{h} \frac{2\pi \tau_\varphi}{h} k_B T \big(\frac{\Delta}{k_B T}-\log (e^{\frac{\Delta}{k_B T}}-1)\big) \notag \\ 
    \rho_{\psi;xx}&=  \frac{m_\psi}{n_\psi \tau_\psi e^2} \notag \\
    \rho_{\psi;xy}&=-\frac{B_{\mathrm{eff}}}{n_\psi e}
\end{align}
where $n_\psi=\frac{1}{m}-n_\varphi$ and $B_{\mathrm{eff}}= n_\varphi$.    We still define $\frac{4\pi^2 \hbar^2}{2 m_\psi a^2}=W_\psi$ as the bandwidth of the fermion $\psi$.  Also the effective magnetic field felt by $\psi$ is obtained from $e B_{\mathrm{eff}}=-m h n_\varphi $.  In the end, we have the expressions

\begin{align}
    \sigma_{\varphi}&=\frac{2\pi \tau_\varphi}{h} k_B T \big(\frac{\Delta}{k_B T}-\log (e^{\frac{\Delta}{k_B T}}-1)\big) \notag \\ 
    \rho_{\psi;xx}&=  \frac{h}{\tau_\psi} \frac{1}{2 W_\psi n_\psi} \notag \\
    \rho_{\psi;xy}&=\frac{1}{n_\psi}-m
    \label{eq:transport_calculation}
\end{align}
In the above, $\sigma$ is in units of $\frac{e^2}{h}$ and $\rho$ is in units of $\frac{h}{e^2}$.  $n_\psi=\frac{1}{m}-n_\varphi$ and $n_\varphi$ can be obtained from Eq.~\ref{eq:self_consistent_n_phi}.  For $\rho_{xy}$, we can see that it gradually evolves from $0$ to $-3$ when increasing $n_\varphi$.

\begin{figure}[ht]
    \centering
    \includegraphics[width=0.95\linewidth]{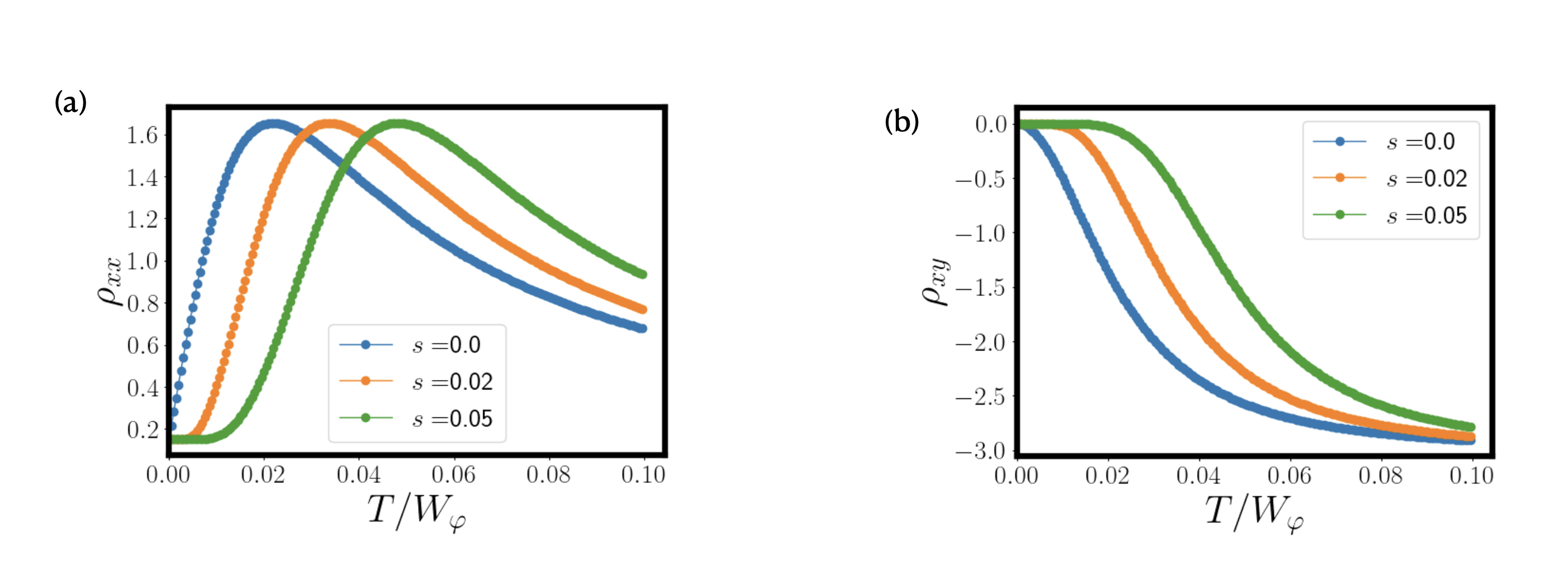}
    \caption{Resistivity $\rho_{xx}$ and $\rho_{xy}$ (in units of $\frac{h}{e^2}$) at the filling $\nu=\frac{1}{3}$, obtained from Eq.~\ref{eq:rho_c} and Eq.~\ref{eq:transport_calculation} using $W_\varphi=W_\psi=1$ and $\frac{h}{\tau_\varphi}=\frac{h}{\tau_\psi}=0.1 W_\varphi$. $s$ is also in units of $W_\varphi$. }
    \label{fig:transport_filling_1_3}
\end{figure}

For simplicity, we set $\frac{h}{\tau_\varphi}=\frac{h}{\tau_\psi}=0.1 W_\psi$ and $W_\psi=W_\varphi$. We plot the results for the filling $\nu=\frac{1}{3}$ with $m=3$ in Fig.~\ref{fig:transport_filling_1_3}. For the filling $\nu=\frac{2}{3}$, we should add an additional integer quantum Hall part to the conductivity, which is shown in the main text. One finds that at small $T$, $\rho_{xx} \sim \sigma_{xx} \sim  \frac{A k_B T}{(A k_B T)^2+(\frac{1}{m})^2}$, where $A \sim \frac{2\pi \tau_\varphi}{h}$.  Interestingly there is a linear T resistivity at low temperature, whose coefficient actually scales with the relaxation time of the critical boson $\varphi$. Here we assume $\tau_\varphi$ is mainly decided by the disorder and thus it is temperature independent. If there is another source of relaxation from interaction which we ignored, there maybe an additional temperature dependence.

\end{document}